\documentstyle[12pt]{article}

\begin{document}
\thispagestyle{empty}
\begin{flushright}
UT-798\\
November, 1997\\
\end{flushright}

\bigskip
\bigskip

\begin{center}
\noindent{\Large \bf Fermion Emission from Five-Dimensional \\
\medskip
                     Black Holes}\\
\bigskip
\bigskip
\noindent{\large Kazuo Hosomichi\footnote{
                 E-mail: hosomiti@hep-th.phys.s.u-tokyo.ac.jp}}\\
\bigskip
{\it Department of Physics, University of Tokyo, \\
\medskip
Tokyo 113, Japan}
\bigskip
\bigskip
\end{center}
\begin{abstract}
  We calculate semiclassically the emission rate of spin $\frac{1}{2}$
particles from charged, nonrotating black holes in $D=5,N=8$ supergravity.
  The relevant Dirac equation is solved by the same approximation
as in the bosonic case.
  The resulting expression for the emission rate
has a form which is predicted from D-brane effective field theory.
\end{abstract}

\newpage

\section{Introduction}
\setcounter{equation}{0}
  Recently there has been a great development in understanding the physics of
black holes in the context of string theory \cite{9705078}.
  The entropy of black holes,
which is proportional to the area of the event horizon,
is explained in string theory by counting the number of microstates
composed of D-branes and open strings
with the ends attached on them \cite{9602043}.
  The Hawking radiation of nonextremal holes is also described
in string picture as the process of creation of a closed string
by the collision of two open strings on D-branes \cite{9604042}.
  The emission rate can be calculated in string picture \cite{9606185}
and agrees precisely with the result of semiclassical calculation
for the case of scalar emission,
including the correct expression for the greybody factors \cite{9609026}.
  The D-brane description of black holes is good in the near-extremal case
as well as the extremal case, and indeed it is good beyond the range of the
value of couplings in which the D-brane description is a valid approximation.
  This indicates that there should be a universal `effective string theory'
which describes the low-energy dynamics of black holes,
whether or not the coupling is weak \cite{9702015,9705192}.

  In this paper we undertake the semiclassical calculation
of the emission rate for the case of massless spin-$\frac{1}{2}$ 
particles\footnote{
  For four-dimensional black holes,
the emission rate has been calculated in \cite{9707124} 
for massless neutral fermions that have no couplings with the gauge fields.
  We think that it is appropriate to consider fermions in supergravity 
theory and take into account the nonminimal coupling with the $U(1)$ field 
strengths when we make a comparison with D-brane calculations.}.
  The black holes we will consider are those in $D=5,N=8$ supergravity,
which are obtained by aligning D5-branes, D-strings and momenta in
type IIB superstring theory compactified on $T^5$.
  The resulting expression for emission rate agrees with the one
predicted from the effective string theory on the worldsheet of the
bound states of D-strings and D5-branes.
  We can see from this expression that the decay process is
interpreted as the collision of a boson and a fermion on the
worldsheet that creates a fermion in spacetime.
  
\section{The Dirac equation in $D=5,N=8$ supergravity}
\setcounter{equation}{0}
  In this section we investigate the Dirac equation in $D=5,N=8$
supergravity.
  What we want to obtain is the precise form 
of the radial wave equation (\ref{radial-wave-equation}) 
and the conserved flux $f$.

  Let us briefly review the basic property of $D=5,N=8$ supergravity.
  This theory contains $8$ gravitini $(\psi^A)$,
$27$ $U(1)$ vector fields $(A_m^{IJ})$,
$48$ Dirac fields $(\chi^{ABC})$ and $42$ scalars besides the graviton.
  This theory has global $E_6$ and local $USp(8)$ symmetries.
  $8$ gravitini and $48$ Dirac fields belong to the irreducible
representations ${\bf 8}$ and ${\bf 48}$ of $USp(8)$, respectively.
  $27$ vector fields belong to the representation ${\bf 27}$ of $E_6$.
  $42$ scalars parameterize the coset space $E_6/USp(8)$ and these degrees
of freedom are contained in a single quantity, the $27$-bein
$V_{IJ}^{AB}$ relating the representations ${\bf 27}$ of $E_6$ and $USp(8)$.
  Note that there is no dynamical gauge fields corresponding to the local
$USp(8)$.
  The $USp(8)$ connection is made out of the $27$-bein, like the
spin connection which is made out of the vielbein.

  The spin-$\frac{1}{2}$ fields in $D=5,N=8$ supergravity obey the following
Dirac equation \cite{Cremmer}:
\begin{equation}
  \tilde{\Gamma}^mD_m\chi^{ABC}+\frac{3i}{2}
  \left[\tilde{\Gamma}^{mn}F_{mn}^{IJ}V_{IJ}^{[AE}\chi^{BC]D}\Omega_{DE}
   -({\rm Trace})^{ABC}\right]=0, \label{Dirac-eq-in-D=5-N=8}
\label{the-Dirac-equation-in-D=5,N=8supergravity}
\end{equation}
\[  \tilde{\Gamma}^m\equiv\Gamma^a(e^{-1})_a^m\;,\;\;
    \tilde{\Gamma}^{mn}\equiv \tilde{\Gamma}^{[m}\tilde{\Gamma}^{n]}\;,\;\;
    F_{mn}^{IJ}=2\partial_{[m}A_{n]}^{IJ} \]   
where $D_m$ is the $SO(1,4)$ and $USp(8)$-covariant derivative and
$\Omega_{AB}$ is the $USp(8)$-invariant tensor.
  The term $({\rm Trace})^{ABC}$ means the subtraction of the trace part
so that the whole left-hand-side become traceless with respect to the indices
$(ABC)$.

  We shall solve the equation 
(\ref{the-Dirac-equation-in-D=5,N=8supergravity}) 
in the nonrotating, nonextremal, triply-charged black hole background. 
  The form of the Einstein metric is well-known \cite{9603100,9603109,9607235}:
\begin{eqnarray}
  ds^2 &=& -\lambda^{-\frac{2}{3}}hdt^2
   +\lambda^{\frac{1}{3}} \left\{ h^{-1}dr^2+r^2 d\Omega_{(3)}^2 \right\},  \\
   \lambda &=&f_1 f_5 f_n,                                        \nonumber \\
    f_i    &=& 1+\frac{r_0^2}{r^2}\sinh^2 \alpha_i\;\; (i=1,5,n), \nonumber \\
    h      &=& 1-\frac{r_0^2}{r^2}.                               \nonumber 
\label{5dmetric}
\end{eqnarray}

  In the charged black hole background, there are excitations in the $U(1)$
gauge fields and the scalars also.
  To know the form for these fields we need some knowledge about the
relation between the fields in the $E_6$ symmetric theory and
the toroidally-compactified type II theory. 

  We first convert the $USp(8)$ indices into the $SO(5)$ indices.
  From the Type IIB on $T^5$ viewpoint, the $USp(8)$ index originates
from the spinor index of the compactified $SO(5)$ times a factor two, 
the multiplicity of SUSY in ten dimensions. 
  We denote the representation {\bf 8} of $USp(8)$ as
\begin{equation}
  \psi^A=(\psi^\alpha,\psi_\alpha),
\end{equation}
and at the same time we introduce the $SO(5)\; \Gamma$-matrix\footnote{
We use the dotted alphabets $\dot{a},\dot{b},\cdots$ and 
$\dot{p},\dot{q},\cdots$ for tangent and world indices of internal $T^5$.
The Greek indices are used for the spinor indices of internal $SO(5)$.}
   $\Gamma_{\dot{a}}=(\Gamma_{\dot{a}})^\alpha_{\;\beta}$ 
and the charge conjugation matrix 
   $C=(C)_{\alpha\beta},\;C^{-1}=(C^{-1})^{\alpha\beta}$
satisfying the standard identities:
\begin{eqnarray}
  \Gamma_{\dot{a}}&=& (\Gamma_{\dot{a}})^\dagger \nonumber \\
  \left\{ \Gamma_{\dot{a}},\Gamma_{\dot{b}} \right\}
                  &=& 2\delta_{{\dot{a}}{\dot{b}}} \nonumber \\
  C\Gamma_{\dot{a}} C^{-1}&=& \Gamma_{\dot{a}}^T
                             =\Gamma_{\dot{a}}^\ast \nonumber \\
  CC^\dagger&=&-CC^\ast=1. 
\end{eqnarray}
  As can be easily verified, $C$ and $C\Gamma_{\dot{a}}$ are antisymmetric 
while $C\Gamma_{{\dot{a}}{\dot{b}}}$ are symmetric.
  The $USp(8)$ invariant tensor $\Omega^{AB}$ is defined as
\begin{equation}
  \Omega_{AB}=\left(\begin{array}{ll}
              \Omega_{\alpha\beta} & \Omega_\alpha^{\;\;\beta} \\
              \Omega^\alpha_{\;\;\beta} & \Omega^{\alpha\beta}
              \end{array}\right)
             =\left(\begin{array}{cc} 
              0 & -\delta_\alpha^\beta \\ \delta^\alpha_\beta& 0 
                      \end{array}\right).
\end{equation}

  We shall then decompose the irreducible representations of $USp(8)$
into those of $SO(5)$.

  Firstly, the representation {\bf 27} of $USp(8)$ is the second-rank
antisymmetric, pseudoreal traceless tensor $F^{AB}$ that satisfies
\begin{eqnarray}
  F^{AB}&=&-F^{BA},\nonumber \\
  (F^{AB})^\ast&=&\Omega_{AC}\Omega_{BD}F^{CD},\nonumber \\
 \Omega_{AB}F^{AB}&=&0.
\end{eqnarray}
  The quantity $F_{mn}^{AB}\equiv V^{AB}_{IJ}F^{IJ}_{mn}$ transforms as
{\bf 27} of $USp(8)$ and we translate it to the quantity
with $SO(5)$ indices as
\begin{eqnarray}
  F^{AB}_{mn}&=& (F^{\alpha\beta},F^\alpha_{\;\;\beta},
                  F_\alpha^{\;\;\beta},F_{\alpha\beta})_{mn}, \nonumber \\
  F^{\alpha\beta}_{mn}&=& \frac{1}{4}
   \left( J_{mn}(C^{-1})^{\alpha\beta}
   +J_{mn\dot{a}}(\Gamma^{\dot{a}}C^{-1})^{\alpha\beta}\right), \nonumber \\ 
  F_{\alpha\beta mn}&=& -\frac{1}{4}
   \left( J^\ast_{mn}(C)_{\alpha\beta}
   +J^\ast_{mn\dot{a}}(C\Gamma^{\dot{a}})_{\alpha\beta}\right), \nonumber \\ 
  F^\alpha_{\;\;\beta mn} &=&\frac{1}{4}
   \left( F_{mn}^{\dot{a}}(\Gamma_{\dot{a}})^\alpha_{\;\;\beta}
   +\frac{i}{2}F_{mn}^{\dot{a}\dot{b}}
    (\Gamma_{\dot{a}\dot{b}})^\alpha_{\;\;\beta} \right).
\label{decomposition-of-27}
\end{eqnarray}

  Similarly, the spin $\frac{1}{2}$ field $\chi^{ABC}$ belongs to {\bf 48}
of $USp(8)$.
  It is antisymmetric and traceless with respect to the indices $(ABC)$ 
\begin{equation}
  \chi^{ABC}=\chi^{[ABC]}\;\;,\;\;\Omega_{AB}\chi^{ABC}=0,
\end{equation}
and obeys the symplectic Majorana condition
\begin{equation}
  \overline{\chi^{ABC}}=C\cdot
  \Omega_{AA'}\Omega_{BB'}\Omega_{CC'}\chi^{A'B'C'},
\end{equation}
where the bar means the Dirac conjugation and $C$ is the charge conjugation
matrix acting on the external $SO(1,4)$ spinor index.
  Again we decompose them into the irreducible representations of SO(5)
as follows:
\begin{eqnarray}
\chi^{ABC}&=& (\chi^{\alpha\beta\gamma},\;\;
               \chi^{\alpha\beta}_{\;\;\;\;\;\gamma},\;\;
               \chi^{\alpha}_{\;\;\beta\gamma},\;\;
               \chi_{\alpha\beta\gamma}) \nonumber \\
\chi^{\alpha\beta\gamma}&=&
    \left\{ (C^{-1})^{\alpha\beta}(C^{-1})^{\gamma\delta}
            -(\Gamma_{\dot{a}}C^{-1})^{\alpha\beta}
             (\Gamma^{\dot{a}}C^{-1})^{\gamma\delta} \right\} \eta_\delta
                                                        \nonumber \\
\chi^{\alpha\beta}_{\;\;\;\;\gamma}&=&
    (\Gamma_{\dot{a}}C^{-1})^{\alpha\beta}\eta^{\dot{a}}_\gamma
    -(C^{-1})^{\alpha\beta}(\Gamma_{\dot{a}}^T\eta^{\dot{a}})_\gamma  
                                                        \nonumber \\ 
\chi^\alpha_{\;\;\beta\gamma}&=&
    (C\Gamma_{\dot{a}})_{\beta\gamma}\xi^{\dot{a}\alpha}
    -C_{\beta\gamma}(\Gamma_{\dot{a}}\xi^{\dot{a}})^\alpha \nonumber \\
\chi_{\alpha\beta\gamma}&=&
  \left\{ C_{\alpha\beta}C_{\gamma\delta}
         -(C\Gamma_{\dot{a}})_{\alpha\beta}(C\Gamma^{\dot{a}})_{\gamma\delta}
  \right\}\xi^\delta
\label{decomposition-of-48}
\end{eqnarray}
where we used the fact that the quantity
\[ (C^{-1})^{\alpha\beta}(C^{-1})^{\gamma\delta}
            -(\Gamma_{\dot{a}}C^{-1})^{\alpha\beta}
             (\Gamma^{\dot{a}}C^{-1})^{\gamma\delta} \;,\;
   C_{\alpha\beta}C_{\gamma\delta}
         -(C\Gamma_{\dot{a}})_{\alpha\beta}
          (C\Gamma^{\dot{a}})_{\gamma\delta} \]
are completely antisymmetric in indices $(\alpha\beta\gamma\delta)$.
  The spinor fields $\xi^{\dot{a}\alpha}, \eta^{\dot{a}\alpha}$
                and $\xi^\alpha,          \eta^\alpha$
are identified, from the IIB on $T^5$ viewpoint,
as the two gravitini with the internal vector index
and the two dilatini, respectively.

  We shall then investigate the $27$-bein $V^{AB}_{IJ}$.
  The infinitesimal $E_6$ transformation on the representation {\bf 27}
which is not in its $USp(8)$ subgroup is expressed as 
\begin{equation}
  \delta F^{IJ}=\Sigma^{IJ}_{\;\;\;\;KL}F^{KL},
\end{equation}
where $\Sigma^{IJ}_{\;\;\;\;KL}$ has the property
\begin{eqnarray}
  \Sigma^{IJKL}&=&\Sigma^{[IJKL]}, \nonumber \\
  \Sigma_I^{\;\;IKL}&=& 0, \nonumber \\
  (\Sigma^{IJKL})^\ast&=&\Sigma_{IJKL},
\end{eqnarray}
and the indices $I,J,K,L,\cdots$ are raised and lowered by 
$\Omega_{IJ} ,\Omega^{IJ}$.
  Rewriting the components in terms of $SO(5)$ tensors, they become
\begin{eqnarray}
  \Sigma^{\alpha\beta\gamma\delta}&=& s^\ast
  \left\{(C^{-1})^{\alpha\beta}(C^{-1})^{\gamma\delta}
         -(\Gamma_{\dot{a}}C^{-1})^{\alpha\beta}
          (\Gamma^{\dot{a}}C^{-1})^{\gamma\delta}
  \right\} \nonumber\\
  \Sigma_{\alpha\beta\gamma\delta}&=& s
  \left\{(C)_{\alpha\beta}(C)_{\gamma\delta}
         -(C\Gamma_{\dot{a}})_{\alpha\beta}(C\Gamma^{\dot{a}})_{\gamma\delta}
  \right\} \nonumber\\
  \Sigma^{\alpha\beta\gamma}_{\;\;\;\;\;\;\delta}&=& 
  -\left\{(C^{-1})^{\alpha\beta}(C^{-1})^{\gamma\tau}
         -(\Gamma_{\dot{a}}C^{-1})^{\alpha\beta}
          (\Gamma^{\dot{a}}C^{-1})^{\gamma\tau}
   \right\} \nonumber \\
  & & \;\;\;\;\;\;\cdot\frac{i}{2}s^{\ast\,\dot{b}\dot{c}}
             (C\Gamma_{\dot{b}\dot{c}})_{\tau\delta} 
             \nonumber\\
  \Sigma_{\alpha\beta\gamma}^{\;\;\;\;\;\;\delta}&=&
   \left\{(C)_{\alpha\beta}(C)_{\gamma\tau}
         -(C\Gamma_{\dot{a}})_{\alpha\beta}(C\Gamma^{\dot{a}})_{\gamma\tau}   
   \right\}\frac{i}{2}s^{\dot{b}\dot{c}}
           (\Gamma_{\dot{b}\dot{c}}C^{-1})^{\tau\delta} \nonumber\\
  \Sigma^{\alpha\beta}_{\;\;\;\;\gamma\delta} &=&
   it_{\dot{a}}\left\{(C^{-1})^{\alpha\beta}(C\Gamma^{\dot{a}})_{\gamma\delta}
              -(\Gamma^{\dot{a}} C^{-1})^{\alpha\beta}(C)_{\gamma\delta}
       \right\} \nonumber \\
   & &+h^{\dot{a}\dot{b}}(\Gamma_{(\dot{a}}C^{-1})^{\alpha\beta}
                             (C\Gamma_{\dot{b})})_{\gamma\delta}
   -h^{\dot{a}}_{\dot{a}}(C^{-1})^{\alpha\beta}(C)_{\gamma\delta}.
\end{eqnarray}
  Note that the number of parameters is $42$, the correct number of 
scalar fields in $D=5,N=8$ supergravity.
  We can recognize these parameters as the scalar degrees of freedom
in type IIB on $T^5$, namely, 
  $s$ and $s^\ast$ as the dilaton $(\phi)$ and the axion $(\chi)$,
  $s^{\dot{a}\dot{b}}$ and $s^{\ast\,\dot{a}\dot{b}}$
  as the components of NS-NS and R-R 2 forms with two internal 
  indices\footnote{
         We distinguish the R-R 2 form (and the quantities related to it)
         from the NS-NS ones by adding a prime$'$.}
  $(B_{\dot{p}\dot{q}},B'_{\dot{p}\dot{q}})$,
  $t^{\dot{a}}$
  as the components of self-dual 4 form with four internal indices 
  $(C^+_{\dot{p}\dot{q}\dot{r}\dot{s}})$ and
  $h^{\dot{a}\dot{b}}$
  as the f$\ddot{\rm u}$nfbein of the internal five dimensions
  $(E_{\dot{p}}^{\dot{a}})$.

  In the present situation the fields 
      $B_{\dot{p}\dot{q}},B'_{\dot{p}\dot{q}}$ and 
      $C^+_{\dot{p}\dot{q}\dot{r}\dot{s}}$
are not excited and we concentrate on the
$s,s^\ast$ and $h^{ij}$-transformations.   
  After some calculation, the explicit form of the $27$-bein 
$V_{IJ}^{AB}$ is expressed in terms of the relation 
$F_{mn}^{AB}= V^{AB}_{IJ}F^{IJ}_{mn}$:
\begin{eqnarray}
  F_{mn}^{\dot{a}}&=& E^{-1}E_{\dot{p}}^{\dot{a}}F_{mn}^{\dot{p}} \;\;,\;\;
                      ( E\equiv {\rm det}(E_{\dot{p}}^{\dot{a}}))   \\
  F_{mn}^{\dot{a}\dot{b}}&=& 
  E_{\dot{p}}^{\dot{a}}E_{\dot{q}}^{\dot{b}}F_{mn}^{\dot{p}\dot{q}} \\
  \left( \begin{array}{c} J_{mn\dot{a}} \\ J^\ast_{mn\dot{a}} 
         \end{array}\right) &=&
  E_{\dot{a}}^{\dot{p}}
  \exp\left(\begin{array}{cc} & s^\ast \\ s& \end{array}\right)
  \left( \begin{array}{c} H_{mn\dot{p}}+iH'_{mn\dot{p}} \\
               H_{mn\dot{p}}-iH'_{mn\dot{p}} \end{array}\right) \\
  \left( \begin{array}{c} J_{mn} \\ J^\ast_{mn} 
         \end{array}\right) &=&
  E\exp\left(\begin{array}{cc} & -s^\ast \\ -s& \end{array}\right)
  \left( \begin{array}{c} K_{mn}+iK'_{mn} \\
               K_{mn}-iK'_{mn} \end{array}\right). 
\end{eqnarray}
  We identify the fields
\[F_{mn}^{\dot{p}}, F_{mn}^{\dot{p}\dot{q}}, 
 H_{mn\dot{p}}, H'_{mn\dot{p}}, K_{mn}, K'_{mn}\]
with those in toroidally compactified Type IIB theory coupled to
KK, D3, NS1, D1, NS5 and D5 charges.
  Note that $s$ parameterize $SU(1,1)/U(1)$ rather than $SL(2,R)/SO(2)$,
and this is why $(H_{mn\dot{p}},H'_{mn\dot{p}})$ and
$(K_{mn},K'_{mn})$ appear in combinations in the above
expressions.

  Rewriting the action of $D=5,N=8$ supergravity
\begin{equation}
\frac{1}{\sqrt{-g}}{\cal L}
   = R-\frac{1}{2}G_{IJ,KL}F_{mn}^{IJ}F^{mn\;KL}
    + \frac{1}{24}\partial_m G_{IJ,KL}\partial^m G^{IJ,KL}+\cdots, 
\end{equation}
\begin{equation}
G_{IJ,KL}=\Omega_{AB}\Omega_{CD}V_{IJ}^{AB}V_{KL}^{CD}
\end{equation}
in terms of the fields with SO(5) indices, and inserting 
\[s=\frac{\phi}{2}\;\;,\;\;H_{mn\dot{p}}=K_{mn}=0, \]
the action becomes
\begin{eqnarray}
  \frac{1}{\sqrt{-g}}{\cal L}&=&
    R(g)+\frac{1}{4}\left( 
   \partial^m G^{\dot{p}\dot{q}}\partial_m G_{\dot{p}\dot{q}}
   +\partial^m G\partial_m G^{-1} \right)
   -\frac{1}{2}\partial^m\phi\partial_m\phi  \nonumber \\
   & & -\frac{1}{4}\left\{
   GG_{\dot{p}\dot{q}}F_{mn}^{\dot{p}}F^{mn\dot{q}}
  +e^{\phi}G^{\dot{p}\dot{q}}H'_{mn\dot{p}}H'^{mn}_{\;\;\;\;\dot{q}}
      \right. \nonumber \\
   & & \;\;\;\;\;\;\;\;\left.+e^{-\phi}GK'_{mn}K'^{mn}\right\}+\cdots,
\label{the-action-in-D=5,N=8}
\end{eqnarray}
\[    G_{\dot{p}\dot{q}}=\delta_{\dot{a}\dot{b}}
       E_{\dot{p}}^{\dot{a}}E_{\dot{q}}^{\dot{b}}\;\;,\;\; 
      G={\rm det}G_{\dot{p}\dot{q}}.  \]

  On the contrary, ten-dimensional type IIB action can be written
(if we drop the four form potential) as\footnote{
  Here the hatted vector indices run $0,1,\cdots,9$,
  while the dotted and undotted indices run $5,\cdots,9$ and $0,1,\cdots,4$,
  respectively.}
\begin{eqnarray}
  S&=&\int d^{10}x\sqrt{-g}\left\{
      R(g)+\frac{1}{4}tr(\partial^m M^{-1}\partial_m M) \right. \nonumber \\
   & & \left. \;\;\;\;\;\; \;\;\;\;\;\;\;\;\;\;\;\;
        -\frac{1}{12}(H\;\;H')M^{-1}
    \left(\begin{array}{c} H\\ H' \end{array}\right)+\cdots\right\},
\end{eqnarray}
\begin{equation}
   M    =e^\phi\left( \begin{array}{cc} 
       1 & \chi \\ \chi & \chi^2+e^{-2\phi} \end{array}\right),
\end{equation}
\[   H_{\hat{m}\hat{n}\hat{p}}=
     3\partial_{[\hat{m}}B_{\hat{n}\hat{p}]}\;,\;
     H'_{\hat{m}\hat{n}\hat{p}}=
     3\partial_{[\hat{m}}B'_{\hat{n}\hat{p}]}.  \]
  This action can be brought to the same form as (\ref{the-action-in-D=5,N=8})
by applying the following procedure:
  we first dimensionally reduce it to five dimensions, that is,
ignore the $x_{5,\cdots,9}$ dependence and insert
      \begin{equation}
      e_{\hat{m}}^{\hat{a}}=\left(\begin{array}{cc}
      \tilde{e}_m^{\;\;a} & A_m^{\dot{m}}\tilde{E}_{\dot{m}}^{\dot{a}} \\
      0 & \tilde{E}_{\dot{m}}^{\dot{a}} \end{array}\right), 
      \end{equation}
      \begin{equation}
      \tilde{g}_{mn}= \tilde{e}_m^{\;\;a}\tilde{e}_n^{\;\;b}\eta_{ab}\;,\;
      \tilde{G}_{\dot{p}\dot{q}}
                    = \tilde{E}_{\dot{p}}^{\;\;\dot{a}}
               \tilde{E}_{\dot{q}}^{\;\;\dot{b}}\delta_{\dot{a}\dot{b}}\;,\;
      \varphi       =-\frac{1}{2}\ln\det \tilde{G}_{\dot{p}\dot{q}}.
      \end{equation}
  We then dualize the three-form field strengths in five dimensions
      \begin{eqnarray}
      \left(\begin{array}{c}H_{mnp} \\ H'_{mnp} \end{array}\right)
        &=& \frac{1}{2\sqrt{-g}}e^\varphi M\varepsilon_{mnpqr}
      \left(\begin{array}{c}K^{qr} \\ K'^{qr}\end{array}\right) \\
      K_{mn}  &=&2\partial_{[m}C_{n]} \\
      K'_{mn} &=&2\partial_{[m}C'_{n]},
      \end{eqnarray}
and Weyl-rescale the external and internal metrics
      \begin{equation}
      \tilde{g}_{mn}=  e^{\frac{2}{3}\varphi}g_{mn}\;\;,\;\;
      \tilde{G}_{\dot{p}\dot{q}}
                    =  e^{-\frac{2}{3}\varphi}G_{\dot{p}\dot{q}}.
      \end{equation}
  Inserting $\chi=K_{mn}=H_{mn\dot{p}}=0$ into the action thus obtained,
we arrive at the same expression as (\ref{the-action-in-D=5,N=8}).

  From the correspondence explained above, we know the form of all the 
bosonic fields in $E_6$ symmetric theory.
  The fields other than metric take the following form:
\begin{eqnarray}
  e^{-2(\phi-\phi_\infty)}&=& f_1^{-1}f_5 \nonumber \\
  G_{\dot{5}\dot{5}}=\cdots= 
  G_{\dot{8}\dot{8}}&=&   f_1^{\frac{1}{6}}f_5^{\frac{1}{6}}f_n^{-\frac{1}{3}}
                                                               \nonumber \\
  G_{\dot{9}\dot{9}}&=& f_1^{-\frac{5}{6}}f_5^{\frac{1}{6}}f_n^{\frac{2}{3}}
                                                               \nonumber \\
{\rm det}G_{\dot{p}\dot{q}}
                    &=& f_1^{-\frac{1}{6}}f_5^{\frac{5}{6}}f_n^{-\frac{2}{3}},
\label{5dscalars}
\end{eqnarray}
\begin{eqnarray}
(F_{mn}^{\dot{p}}\;,\;H'_{mn\dot{p}}\;,\;K'_{mn})
    &=& (2\partial_{[m}A_{n]}^{\dot{p}}\;,\;
         2\partial_{[m}B'_{n]\dot{p}}\;,\;
         2\partial_{[m}C'_{n]}) \nonumber \\  
  A_0^{\;\;\dot{9}}&=& \frac{r_0^2\sinh\alpha_n\cosh\alpha_n}
                      {r^2+r_n^2}                              \nonumber \\
  B'_{0\dot{9}}    &=& \frac{r_0^2\sinh\alpha_1\cosh\alpha_1}
                      {r^2+r_1^2}                              \nonumber \\
  C'_0        &=& \frac{r_0^2\sinh\alpha_5\cosh\alpha_5}
                      {r^2+r_5^2}.
\label{5dvectors}
\end{eqnarray}
  Putting all these into (\ref{the-Dirac-equation-in-D=5,N=8supergravity}),
we obtain the Dirac equation for the fields
\[ \xi^{\dot{a}\alpha}, \xi^\alpha,\eta^{\dot{a}\alpha},\eta^\alpha .\]
  It is rather easy to see that the $USp(8)$ connection in the covariant 
derivative vanishes, but calculating the interaction term is quite tedious. 
  After some calculation, we have found that the interaction term is 
diagonal only for a subset of the spinor fields.
  Define 
\begin{equation}
 \psi^{\dot{b}\alpha}_{(\pm)}\equiv 
  \xi^{\dot{b}\alpha}\pm i\eta^{\dot{b}\alpha}
  -\frac{1}{4}\sum_{a=5,\cdots,8}
    (\Gamma^{\dot{b}}\Gamma _{\dot{a}})^\alpha_{\;\;\beta}
    (\xi^{\dot{a}\beta}\pm i\eta^{\dot{a}\beta}),
\end{equation}
then the Dirac equation for these fields becomes
\begin{eqnarray}
 \tilde{\Gamma}^mD_m \psi^{\dot{b}\alpha}_{(\pm)}
  +\frac{i}{8}\tilde{\Gamma}^{mn}\left\{
 \mp e^{-\frac{\phi}{2}}\sqrt{G}K'_{mn}\psi^{\dot{b}\alpha}_{(\pm)}
                          \right.
   \;\;\;\;\;\;\;\;\;\;\;\;\;\;\;\;\;\;\;\;\;\;\; & & \nonumber \\
 \left. \pm e^{\frac{\phi}{2}}\sqrt{G^{\dot{9}\dot{9}}}H'_{mn\dot{9}}
      (\Gamma^{\dot{9}})^\alpha_{\;\;\beta}\psi^{\dot{b}\beta}_{(\pm)}
  -  \sqrt{G^{-1}G_{\dot{9}\dot{9}}}F_{mn}^{\dot{9}}
      (\Gamma^{\dot{9}})^\alpha_{\;\;\beta}\psi^{\dot{b}\beta}_{(\pm)}\right\}
 &=&0,
\end{eqnarray}
or, denoting $\psi^{\dot{a}\alpha}_{(\pm)}$ generally as $\psi$ and 
letting $\epsilon_1=\pm 1$ for $\psi_{(\pm)}$ and $\epsilon_2$ 
the eigenvalue of $\Gamma_{\dot{9}}$, this equation can be rewritten as
\begin{eqnarray}
 \left[\tilde{\Gamma}^mD_m
  +\frac{i}{8}\tilde{\Gamma}^{mn}\left\{
  -\epsilon_1 e^{-\frac{\phi}{2}}\sqrt{G}K'_{mn}\right.\right. 
  \;\;\;\;\;\;\;\;\;\;\;\;\;\;\;\;\;\;\;\;\;\;\;\;\;\;\;\; & & \nonumber \\
   \left.\left.
  +\epsilon_1\epsilon_2 e^{\frac{\phi}{2}}
               \sqrt{G^{\dot{9}\dot{9}}}H'_{mn\dot{9}}
  - \epsilon_2 \sqrt{G^{-1}G_{\dot{9}\dot{9}}}F_{mn}^{\dot{9}}\right\}
 \right]\psi& =&0.
\label{the-Dirac-equation-diagonalized}
\end{eqnarray}
  Note that these fields are interpreted, from the type IIB on $T^5$
viewpoint, as the gravitini whose vector indices are those of internal $T^4$.

  In the present case where three gauge fields are excited, there
are three contributions in the interaction term.
  For the other subset of the spinor fields the three contributions cannot be 
diagonalized simultaneously and the Dirac equation for them will be
much more difficult to solve.

  In the following we shall consider (\ref{the-Dirac-equation-diagonalized})
with $\epsilon_1=\epsilon_2=-1$.
  Putting (\ref{5dmetric}), (\ref{5dscalars}) and (\ref{5dvectors})
into (\ref{the-Dirac-equation-diagonalized}), it becomes as
\begin{equation}
  -r\lambda^{\frac{1}{2}}h^{-\frac{1}{2}}\Gamma^0\partial_0\psi
   =  \left\{ r\Gamma^i\partial_i+\left(
              (h^{\frac{1}{2}}-1)x^i\partial_i+C+i\Delta\Gamma^0\right)
              \right\}\Pi\psi 
  \label{Dirac-eq-in-bh-background}
\end{equation}
where the index $i$ runs $1,\cdots,4$ and $C$ is a function of $r$ and
\begin{eqnarray}
  \Delta&=& \frac{r_0^2}{4r^2}\left( f_1^{-1}\sinh 2\alpha_1
             +f_5^{-1}\sinh 2\alpha_5+f_n^{-1}\sinh 2\alpha_n \right),\\
  \Pi   &=& \frac{\Gamma^i x_i}{r}
\end{eqnarray}       
  Note that we can change the function $C$ arbitrarily by rescaling $\psi$.

  Let us see the rotational invariance of (\ref{Dirac-eq-in-bh-background}).
  Defining the orbital angular momentum operator $L_{ij}$ and 
the spin operator $\Sigma_{ij}$ as
\begin{eqnarray}
  L_{ij}      &=& -i(x_i\partial_j-x_j\partial_i) \nonumber \\
  \Sigma_{ij} &=& -\frac{i}{4}[\Gamma_i,\Gamma_j] ,
\end{eqnarray}
the Hamiltonian in (\ref{Dirac-eq-in-bh-background}) commutes with 
the total angular momentum $J_{ij}=L_{ij}+\Sigma_{ij}$ . 
  Therefore we shall obtain the solution as an eigenfunction of 
$ J^2\equiv J_{ij}J_{ij}$ and, say, $J_{12} ,J_{34}$.
  The solution can certainly be expressed as a linear combination of 
eigenfunctions of $\left\{L^2,J^2,J_{12},J_{34}\right\}$,
or, two spins $(j,j')$ of $SU(2)\times SU(2)\simeq SO(4)$ and their
$z$-components $(m,m')$ besides the orbital angular momentum $l$.  
  We denote them as $\Psi^{(l)}_{jj'mm'}$.
  The basic properties of $\Psi^{(l)}_{jj'mm'}$ are listed in the appendix.

  Rewriting the Dirac equation (\ref{Dirac-eq-in-bh-background})
with the operators $(\Sigma\cdot L)$ and $\Pi$, it becomes, 
after a suitable rescaling, as
\begin{equation}
  -r\lambda^{\frac{1}{2}}h^{-\frac{1}{2}}\Gamma^0\partial_0\psi=
  \left[rh^{\frac{1}{2}}\frac{d}{dr}
        +(\Sigma\cdot L)+\frac{3}{2}+i\Delta\Gamma^0 \right]\Pi\psi. 
  \label{Dirac-eq-simplified}
\end{equation}
  Putting the form of the general solution
\begin{equation}
  \psi=e^{-i\omega t}\cdot\left\{
       F_{l\pm 1,l}\Psi^{(l)}_{\frac{l\pm 1}{2}\frac{l}{2}mm'} 
      +G_{l\pm 1,l}\Psi^{(l\pm 1)}_{\frac{l\pm 1}{2}\frac{l}{2}mm'} \right\}
\end{equation}
into (\ref{Dirac-eq-simplified}), we obtain the radial wave equation
\begin{eqnarray}
 \omega r\lambda^{\frac{1}{2}}h^{-\frac{1}{2}}F_{l+1,l}
  &=&\left[rh^{\frac{1}{2}}\frac{d}{dr}+(l+\frac{3}{2}+\Delta)\right]
     G_{l+1,l} \nonumber \\
-\omega r\lambda^{\frac{1}{2}}h^{-\frac{1}{2}}G_{l+1,l}
  &=&\left[rh^{\frac{1}{2}}\frac{d}{dr}-(l+\frac{3}{2}+\Delta)\right]
     F_{l+1,l}. 
   \label{eq-for-F-and-G}
\label{radial-wave-equation}
\end{eqnarray}
  We can derive the equations for $F_{l,l+1},G_{l,l+1}$ from the above 
equations by substituting $\left\{ F_{l,l+1},G_{l,l+1}\right\}$ 
for $\left\{-G_{l+1,l},F_{l+1,l}\right\}$ and $-\Delta$ for $\Delta$. 

  Finally, for spinor fields satisfying (\ref{Dirac-eq-simplified}),
the following equation holds:
\[ \partial_0 \left(\int d\Omega_{(3)}
             \lambda^{\frac{1}{2}}h^{-1}\bar{\psi}\Gamma_0\psi\right)
 = -\frac{d}{dr}\left(\int d\Omega_{(3)}\bar{\psi}\Pi\psi\right). \]
So we define the conserved flux as
$f\equiv \int d\Omega_{(3)}\bar{\psi}\Pi\psi$.
  $f$ is expressed by the radial wave functions $F_{l+1,l},G_{l+1,l}$ as
\begin{equation}
  f\equiv i(G_{l+1,l}^\ast F_{\l+1,l}-F_{l+1,l}^\ast G_{\l+1,l}).
\end{equation}

\section{The solution}
\setcounter{equation}{0}
  Let us solve the radial wave equation (\ref{radial-wave-equation}).
  As in the case of scalar field, the equation cannot be solved analytically
and what we shall do in the following is to divide the range of $r$
in two (or more) regions and obtain the approximate solution  
in each region and then continue them.
  We also limit ourselves to the `dilute gas approximation' 
or the case in which two of the three boost parameters, namely,
$\alpha_1$ and $\alpha_5$, are large.

  To begin with, we define some radial variables for convenience:
\begin{equation}
  \frac{r^2}{r_0^2}=\frac{1}{2}(1+\cosh 2\rho)\;,\;
  y=\tanh\rho\;,\;
  x=1-y\;,
\end{equation}
and we recall the relation between the boost parameters and the mass
and temperature of the black hole\cite{9705192}:
\begin{eqnarray}
  \frac{1}{2\pi T_R}&=&\frac{1}{2\pi T_H}+\frac{1}{2\pi T_-}\nonumber \\
  \frac{1}{2\pi T_L}&=&\frac{1}{2\pi T_H}-\frac{1}{2\pi T_-}\nonumber \\
  \frac{1}{2\pi T_H}&=& r_0\cosh\alpha_1\cosh\alpha_5\cosh\alpha_n\nonumber \\ 
  \frac{1}{2\pi T_-}&=& r_0|\sinh\alpha_1\sinh\alpha_5\sinh\alpha_n|
                                                                  \nonumber \\
  M                 &=&\frac{r_0^2}{2}
                       (\cosh 2\alpha_1+\cosh 2\alpha_5+\cosh 2\alpha_n)
\end{eqnarray} 
where the temperatures $T_H,T_-$ are defined by 
$(\frac{1}{2\pi})$ times the surface accelerations
of the outer and inner event horizons.
  In the dilute gas approximation the temperatures can be written as
\begin{eqnarray}
T_H^{-1}=\frac{\pi
r_0}{2}e^{|\alpha_1|+|\alpha_5|}\cdot\cosh\alpha_n&,&
T_R^{-1}=\frac{\pi
r_0}{2}e^{|\alpha_1|+|\alpha_5|+|\alpha_n|}\nonumber \\
T_-^{-1}=\frac{\pi
r_0}{2}e^{|\alpha_1|+|\alpha_5|}\cdot\sinh|\alpha_n|&,&
T_L^{-1}=\frac{\pi
r_0}{2}e^{|\alpha_1|+|\alpha_5|-|\alpha_n|}
\end{eqnarray}

  In terms of the new variable $y$, the functions $h,\lambda,\Delta$ are
expressed as
\begin{eqnarray}
  h        &=& y^2 \nonumber \\
  \lambda  &=& \frac{1}{2\pi r_0 T_H}(1-y^2\tanh^2\alpha_1)
               (1-y^2\tanh^2\alpha_5)(1-y^2\tanh^2\alpha_n) \nonumber \\
  \Delta   &=& \frac{1-y^2}{4}\frac{d}{dy}\ln\left[
               \frac{1+y\tanh\alpha_1}{1-y\tanh\alpha_1}
               \frac{1+y\tanh\alpha_5}{1-y\tanh\alpha_5}
               \frac{1+y\tanh\alpha_n}{1-y\tanh\alpha_n}\right].
\end{eqnarray}

  We define the new set of radial wave functions $X_1,X_2$ as
\begin{eqnarray}
  X_1&=& (V_2/V_1)^{\frac{1}{4}}F_{l+1,l}, \nonumber \\
  X_2&=& (V_1/V_2)^{\frac{1}{4}}G_{l+1,l},           \\
  V_1&=& (\frac{\omega}{2\pi T_H})y^{-1}(1-y)^{q-3}(1+y)^{-q} \nonumber \\
     & &\cdot(1+y\tanh\alpha_1)(1+y\tanh\alpha_5)(1+y\tanh\alpha_n)\nonumber\\ 
  V_2&=& (\frac{\omega}{2\pi T_H})y^{-1}(1+y)^{q-3}(1-y)^{-q} \nonumber \\
     & &\cdot(1-y\tanh\alpha_1)(1-y\tanh\alpha_5)(1-y\tanh\alpha_n).\nonumber
\end{eqnarray}
where $q$ is an arbitrary constant.
  Then the equations for $X_1,X_2$ are
\begin{eqnarray}
  \left[\frac{d}{dy}+\frac{l+q}{1-y^2}\right]X_2&=& V_1X_1 \nonumber \\
  \left[\frac{d}{dy}-\frac{l+q}{1-y^2}\right]X_1&=&-V_2X_2.
\label{new-radial-equation}
\end{eqnarray}

\subsection{The horizon region}
  We first solve the equation in `the horizon region' or the region of $y$
in which 
  \[ |\coth\alpha_{1,5}|-1 \ll (1-y)\]
or
  \[ r\ll r_0\sinh \alpha_{1,5}. \]
  In this region it is a good approximation to set 
\[\tanh\alpha_1={\rm sgn}\alpha_1\;,\;
  \tanh\alpha_5={\rm sgn}\alpha_5.\]
  Then the equation (\ref{new-radial-equation}) with 
\[q=\theta(\alpha_1)+\theta(\alpha_5)+\theta(\alpha_n)\]
becomes, under the dilute gas approximation,
\begin{eqnarray}
  \left[(1-y^2)\frac{d}{dy}+(l+q)\right]X_2&=&
  y^{-1}(1-y\varsigma)
  \left(\frac{          \omega}{2\pi T_H}+
        \frac{y\varsigma\omega}{2\pi T_-}\right)X_1 \nonumber \\
  \left[(1-y^2)\frac{d}{dy}-(l+q)\right]X_1&=&-
  y^{-1}(1+y\varsigma)
  \left(\frac{          \omega}{2\pi T_H}-
        \frac{y\varsigma\omega}{2\pi T_-}\right)X_2,\\
  \varsigma&=& {\rm sgn}\alpha_n \nonumber 
  \label{equation-for-X1,X2-in-the-horizon-region}
\end{eqnarray}
or, defining $X_{\pm}\equiv X_1\pm iX_2$, the equation for them becomes
\begin{eqnarray}
\left[ (1-y^2)\frac{d}{dy}
      -i\left\{\frac{\omega}{2\pi T_H y}-\frac{\omega y}{2\pi T_-}\right\}
\right]X_+&=& \left(l+q-\frac{i\varsigma\omega}{2\pi T_L}\right)X_-,
                             \nonumber \\
\left[ (1-y^2)\frac{d}{dy}
      +i\left\{\frac{\omega}{2\pi T_H y}-\frac{\omega y}{2\pi T_-}\right\}
\right]X_-&=& \left(l+q+\frac{i\varsigma\omega}{2\pi T_L}\right)X_+.
  \label{equation-for-X+,X--in-the-horizon-region}
 \end{eqnarray}
  The equation (\ref{equation-for-X+,X--in-the-horizon-region}) can be solved
without further approximation.
  The solution with the appropriate boundary condition is as follows:
\begin{eqnarray}
  X_+&=& y^{1-\frac{i\omega}{2\pi T_H}}(1-y^2)^{\frac{l+q}{2}}
         F(\alpha+1,\beta,\gamma+1;y^2)\sqrt{2}\gamma^{-1}
         \left(\frac{l+q}{2}-\frac{i\omega\varsigma}{4\pi T_L}\right),
         \nonumber \\
  X_-&=& y^{-\frac{i\omega}{2\pi T_H}}(1-y^2)^{\frac{l+q}{2}}
         F(\alpha,\beta,\gamma;y^2)\sqrt{2}, 
\end{eqnarray}
\begin{equation}
\alpha=\frac{l+q}{2}  -\frac{i\omega}{4\pi T_L} \;\;,\;\;
\beta =\frac{l+q+1}{2}-\frac{i\omega}{4\pi T_R} \;\;,\;\;
\gamma=\frac{1}{2}-    \frac{i\omega}{2\pi T_H}.
\end{equation}
  The normalization of the solution is such that the absorbing flux
at the horizon is $f_{\rm abs}=1$.

  From the solution obtained above we can derive the leading order behavior
of $F_{l+1,l},G_{l+1,l}$ at small $(1-y)$. 
  For $\alpha_n>0$, 
\begin{eqnarray}
  F_{l+1,l}&\sim&(\frac{T_L}{T_R})^{\frac{1}{4}}\left[
     (1-y^2)^{-\frac{1}{2}(l+q_0+\frac{1}{2})}
     \frac{\Gamma(\frac{1}{2}-\frac{i\omega}{2\pi T_H})\Gamma(l+q_0+1)}
          {\Gamma(\frac{l+q_0+1}{2}-\frac{i\omega}{4\pi T_L})
           \Gamma(\frac{l+q_0+2}{2}-\frac{i\omega}{4\pi T_R})} 
                              \right. \nonumber \\
& & \left.+(1-y^2)^{\frac{1}{2}(l+q_0+\frac{3}{2})}
     \frac{\frac{-i\omega}{4\pi T_L}
           \Gamma(\frac{1}{2}-\frac{i\omega}{2\pi T_H})\Gamma(-l-q_0-1)}
          {\Gamma(\frac{-l-q_0+1}{2}-\frac{i\omega}{4\pi T_L})
           \Gamma(\frac{-l-q_0}{2}-\frac{i\omega}{4\pi T_R})}\right] \\
  G_{l+1,l}&\sim&(\frac{T_R}{T_L})^{\frac{1}{4}}\left[
     (1-y^2)^{-\frac{1}{2}(l+q_0-\frac{1}{2})}
     \frac{\frac{\omega}{4\pi T_R}
           \Gamma(\frac{1}{2}-\frac{i\omega}{2\pi T_H})\Gamma(l+q_0)}
          {\Gamma(\frac{l+q_0+1}{2}-\frac{i\omega}{4\pi T_L})
           \Gamma(\frac{l+q_0+2}{2}-\frac{i\omega}{4\pi T_R})}
                              \right. \nonumber \\
& & \left.+(1-y^2)^{\frac{1}{2}(l+q_0+\frac{1}{2})}
     \frac{i\Gamma(\frac{1}{2}-\frac{i\omega}{2\pi T_H})\Gamma(-l-q_0)}
          { \Gamma(\frac{-l-q_0+1}{2}-\frac{i\omega}{4\pi T_L})
            \Gamma(\frac{-l-q_0}{2}-\frac{i\omega}{4\pi T_R})}\right],\\
q_0&=& \theta(\alpha_1)+\theta(\alpha_5)
\end{eqnarray}
and the expression for $\alpha_n<0$ is obtained by $T_L\leftrightarrow T_R$.

\medskip

\noindent{\large \bf String symmetry}

  Let us comment on the `effective string symmetry' of 
(\ref{equation-for-X+,X--in-the-horizon-region}).
  As was discussed in \cite{9705192} for the case of minimally coupled scalars,
the approximate wave equation near the horizon should have a 
$SL(2,R)\times SL(2,R)$ symmetry if the dynamics and the microstructure
of black holes are well described by the `effective string' theory.
  This also holds in the present case.
  Define a two-component function 
$\Phi(\rho,\tau,\sigma)$ as
\begin{eqnarray}
  \Phi  &=& (\phi_1,\phi_2)^T, \\
  \phi_1&=& e^{-\frac{i}{2}(\tau-\sigma)-\frac{i\pi}{4}
               +\frac{\omega\tau}{2\pi T_H}+\frac{\omega\sigma}{2\pi T_-}}
            (\sinh 2\rho)^{-\frac{1}{2}}X_-(\rho)\varsigma,\nonumber \\
  \phi_2&=& e^{ \frac{i}{2}(\tau-\sigma)+\frac{i\pi}{4}
               +\frac{\omega\tau}{2\pi T_H}+\frac{\omega\sigma}{2\pi T_-}}
            (\sinh 2\rho)^{-\frac{1}{2}}X_+(\rho),
\end{eqnarray}
and the $SL(2,R)$ generators as
\begin{eqnarray}
  (\Sigma_1,\Sigma_2,\Sigma_3)&=& 
  (\frac{\sigma_1}{2},\frac{\sigma_2}{2},\frac{i\sigma_3}{2}), \\
  R_1&=&\frac{1}{2}\sin(\tau+\sigma)\frac{\partial}{\partial\rho}
        +\frac{1}{2}\cos(\tau+\sigma)
      \left(\coth\rho\frac{\partial}{\partial\tau}
           +\tanh\rho\frac{\partial}{\partial\sigma}\right), \nonumber \\
  R_2&=&-\frac{1}{2}\cos(\tau+\sigma)\frac{\partial}{\partial\rho}
        +\frac{1}{2}\sin(\tau+\sigma)
      \left(\coth\rho\frac{\partial}{\partial\tau}
           +\tanh\rho\frac{\partial}{\partial\sigma}\right), \nonumber \\
  R_3&=&\frac{1}{2}\left(\frac{\partial}{\partial\tau}
                        +\frac{\partial}{\partial\sigma}\right), \\
  L_{1,2,3}&=& R_{1,2,3}\;\;\mbox{with}\;\sigma\;\mbox{replaced by}\;-\sigma, 
\end{eqnarray}
then the equation for $\Phi$ can be written as
\begin{equation}
  2(L\cdot \Sigma)\Phi\equiv 2(-L_1\Sigma_1-L_2\Sigma_2+L_3\Sigma_3)\Phi
  =\frac{1+\varsigma(l+q)}{2}\Phi.
\end{equation}

\subsection{The asymptotic region}
  Next we shall solve the equation (\ref{radial-wave-equation}) 
in the `asymptotic region', where
\[ x=(1-y) \ll |\coth\alpha_{1,5}|-1 \]
or
\[ r \gg r_0\sinh\alpha_{1,5}. \]
  In this region we shall treat (\ref{radial-wave-equation}) with $q=-l$.
  (\ref{radial-wave-equation}) becomes as
\begin{equation}
  \frac{d}{dy}X_2= V_1X_1\;,\;
  \frac{d}{dy}X_1=-V_2X_2
\end{equation}  
\begin{eqnarray}
  V_1&=& (\frac{\omega}{2\pi T_H})y^{-1}(1-y)^{-l-3}(1+y)^{l} \nonumber \\
     & &\cdot(1+y\tanh\alpha_1)(1+y\tanh\alpha_5)(1+y\tanh\alpha_n)\nonumber\\ 
  V_2&=& (\frac{\omega}{2\pi T_H})y^{-1}(1+y)^{-l-3}(1-y)^{l} \nonumber \\
     & &\cdot(1-y\tanh\alpha_1)(1-y\tanh\alpha_5)(1-y\tanh\alpha_n).
\end{eqnarray}
and the second-order differential equation can be easily derived:
\begin{eqnarray}
\left[ \frac{d^2}{dy^2}+V_1V_2+\left(\frac{V_2'}{2V_2}\right)'
       -\left(\frac{V_2'}{2V_2}\right)^2\right](V_2^{-\frac{1}{2}}X_1)&=&0,
\label{differential-equations-for-X1}   \\
\left[ \frac{d^2}{dy^2}+V_1V_2+\left(\frac{V_1'}{2V_1}\right)'
       -\left(\frac{V_1'}{2V_1}\right)^2\right](V_1^{-\frac{1}{2}}X_2)&=&0.
\label{differential-equations-for-X2}
\end{eqnarray}
where the prime represents the derivative with respect to $y$.
  Leaving only the leading two terms in the $x$-expansion, these become as
\begin{eqnarray}
\left[ \frac{d^2}{dx^2}+\frac{r_0^2\omega^2}{x^3}
      +\frac{M\omega^2+1-(l+1)^2}{4x^2}\right](V_2^{-\frac{1}{2}}X_1)&=&0,
\label{equation-in-the-asymptotic-region-for-X1}            \\
\left[ \frac{d^2}{dx^2}+\frac{r_0^2\omega^2}{x^3}
      +\frac{M\omega^2+1-(l+2)^2}{4x^2}\right](V_1^{-\frac{1}{2}}X_2)&=&0,
\label{equation-in-the-asymptotic-region-for-X2}
\end{eqnarray}
and the solution is expressed in terms of the Bessel functions as
\begin{eqnarray}
\xi&\equiv& r_0\omega/\sqrt{2x},\nonumber \\
X_1&     =& V_2^{\frac{1}{2}}\xi^{-1}
          \left(A J_{  l+1 -2\varepsilon(l+2)}(\xi)+
                B J_{-(l+1)+2\varepsilon(l+2)}(\xi)\right), \nonumber \\
X_2&     =& V_1^{\frac{1}{2}}\xi^{-1}
          \left(A J_{  l+2 -2\varepsilon(l+1)}(\xi)-
                B J_{-(l+2)+2\varepsilon(l+1)}(\xi)\right),\\
F_{l+1,l}&=& (V_1V_2)^{\frac{1}{4}}\xi^{-1}
          \left(A J_{  l+1 -2\varepsilon(l+2)}(\xi)+
                B J_{-(l+1)+2\varepsilon(l+2)}(\xi)\right), \nonumber \\
G_{l+1,l}&=& (V_1V_2)^{\frac{1}{4}}\xi^{-1}
          \left(A J_{  l+2 -2\varepsilon(l+1)}(\xi)-
                B J_{-(l+2)+2\varepsilon(l+1)}(\xi)\right),
                                  \label{solution-in-the-asymptotic-region} \\
\varepsilon&=& \frac{M\omega^2}{4(l+1)(l+2)}.
\end{eqnarray}
  From the above solution we can calculate the incoming flux at infinity
\begin{equation}
f_{\rm in}=\frac{1}{\pi r_0^2\omega^2}|A-(-)^l B|^2,
\end{equation} 
and the behavior of $F_{l+1,l},G_{l+1,l}$ at small $\xi$
\begin{eqnarray}
  F_{l+1,l}&\simeq& (r_0\omega)^{-1}\left[
       \frac{A2^{-l-1}}{\Gamma(l+2)}\xi^{l+\frac{3}{2}}
      +\frac{B2^{l+1}}{\Gamma(-l)}\xi^{-l-\frac{1}{2}}\right]\nonumber\\   
  G_{l+1,l}&\simeq& (r_0\omega)^{-1}\left[
       \frac{A2^{-l-2}}{\Gamma(l+3)}\xi^{l+\frac{5}{2}}
      -\frac{B2^{l+2}}{\Gamma(-l-1)}\xi^{-l-\frac{3}{2}}\right].
\end{eqnarray}

\subsection{Matching the solutions}
  We would like to connect the two solutions in the `matching region'
\[ |\coth\alpha_{1,5}|-1 \ll x \ll 1 .\]
  The first thing to do is to see whether the solution in the asymptotic
region (\ref{solution-in-the-asymptotic-region}) is good in the matching 
region also.
  Actually the differential equations for $X_1$ and $X_2$ in the asymptotic 
region (\ref{differential-equations-for-X1}) and
       (\ref{differential-equations-for-X2}) 
have regular singularities at zeroes or poles of $V_2$ and $V_1$.
  Among others they have zeroes at $x=1-\coth\alpha_{1,5}$ and 
$x=1+\coth\alpha_{1,5}$ respectively and some of them are slightly below $x=0$.
  We have to take into account the singularities arising from these zeroes 
near $x=0$ when we are in the matching region, though we could neglect
them in the asymptotic region.
  The effect of these singularities is different according to the signs of 
$\alpha_{1,5}$.

\subsubsection{$\alpha_1>0\;,\;\alpha_5>0$}
  In this case $V_1$ does not have zeroes near $x=0$
and the differential equation for $X_2$ or
(\ref{differential-equations-for-X2}) has no singularities near $x=0$.
  Therefore the solutions for $X_2$ and $G_{l+1,l}$ in the asymptotic region 
are also good in the matching region and we obtain the coefficients
$A,B$ by matching the solutions for $G_{l+1,l}$ in the horizon region and 
the asymptotic region.
  They become, for $\alpha_n>0$, as
\begin{eqnarray}
  A&=& \frac{\Gamma(\frac{1}{2}-\frac{i\omega}{2\pi T_H})
             \Gamma(l+2)\Gamma(l+3)}
            {\Gamma(\frac{l+3}{2}-\frac{i\omega}{4\pi T_L})
             \Gamma(\frac{l+4}{2}-\frac{i\omega}{4\pi T_R})}
       2^{l+\frac{1}{2}}(r_0\omega)^{-l-1}
       \left(\frac{\omega}{\pi T_R}\right)^{\frac{1}{2}} \nonumber \\
  B&=& \frac{i\Gamma(\frac{1}{2}-\frac{i\omega}{2\pi T_H})
             \Gamma(-l-2)\Gamma(-l-1)}
            {\Gamma(\frac{-l-2}{2}-\frac{i\omega}{4\pi T_R})
             \Gamma(\frac{-l-1}{2}-\frac{i\omega}{4\pi T_L})}
       2^{-l-\frac{3}{2}}(r_0\omega)^{l+3}
       \left(\frac{\omega}{\pi T_R}\right)^{-\frac{1}{2}}
\end{eqnarray}
and $T_R\leftrightarrow T_L$ for $\alpha_n<0$.
  On the contrary, the differential equation for $X_1$ or
(\ref{differential-equations-for-X1}) has additional two regular singularities
at small negative $x$ besides an irregular singularity at $x=0$.
  So the solutions for $F_{l+1,l}$ in the asymptotic region is no longer good
in the matching region and actually there is no simple way to approximate 
$F_{l+1,l}$ in the matching region.
 
\subsubsection{$\alpha_1<0\;,\;\alpha_5<0$}
  This case is similar to the previous subsubsection.
  This time the solution for $F_{l+1,l}$ in the asymptotic region is good also
in the matching region.
  Matching $F_{l+1,l}$ gives, for $\alpha_n>0$,
\begin{eqnarray}
  A&=& \frac{\Gamma(\frac{1}{2}-\frac{i\omega}{2\pi T_H})
             \Gamma(l+1)\Gamma(l+2)}
            {\Gamma(\frac{l+1}{2}-\frac{i\omega}{4\pi T_L})
             \Gamma(\frac{l+2}{2}-\frac{i\omega}{4\pi T_R})}
       2^{l+\frac{3}{2}}(r_0\omega)^{-l}
       \left(\frac{\omega}{\pi T_L}\right)^{-\frac{1}{2}} \nonumber \\
  B&=& \frac{-i\Gamma(\frac{1}{2}-\frac{i\omega}{2\pi T_H})
             \Gamma(-l-1)\Gamma(-l)}
            {\Gamma(\frac{-l}{2}-\frac{i\omega}{4\pi T_R})
             \Gamma(\frac{-l+1}{2}-\frac{i\omega}{4\pi T_L})}
       2^{-l-\frac{5}{2}}(r_0\omega)^{l+2}
       \left(\frac{\omega}{\pi T_L}\right)^{\frac{1}{2}}
\end{eqnarray}
and $T_R\leftrightarrow T_L$ for $\alpha_n<0$.
Approximating $G_{l+1,l}$ in the matching region is difficult because 
of the additional two singularities
near $x=0$ in the differential equation for $X_2$.

\subsubsection{$\alpha_1>0\;,\;\alpha_5<0$}
  In this case both $V_1$ and $V_2$ have a zero near $x=0$. 
  Neither the solution for $F_{l+1,l}$ in the asymptotic region 
nor the one for $G_{l+1,l}$ are good in the matching region.
  So in the matching region we have to consider, instead of 
(\ref{equation-in-the-asymptotic-region-for-X1}) and
(\ref{equation-in-the-asymptotic-region-for-X2}),
the following differential equations:
\begin{eqnarray}
  \left[ \frac{d^2}{dx^2}+\frac{r_0^2\omega^2}{8x^3}
        +\frac{M\omega^2+1-(l+1)^2}{4x^2} 
          \;\;\;\;\;\;\;\;\;\;\;\;\;\;\;\;\right. & & \nonumber \\
  \left.-\frac{3}{4(x+\delta_1)^2}
        -\frac{2l}{4x(x+\delta_1)}\right](V_2^{-\frac{1}{2}}X_1)&=&0,
                                                       \nonumber \\
  \left[ \frac{d^2}{dx^2}+\frac{r_0^2\omega^2}{8x^3}
        +\frac{M\omega^2+1-(l+2)^2}{4x^2} 
          \;\;\;\;\;\;\;\;\;\;\;\;\;\;\;\;\right. & & \nonumber \\
  \left.-\frac{3}{4(x+\delta_5)^2}
        +\frac{2(l+3)}{4x(x+\delta_5)}\right](V_1^{-\frac{1}{2}}X_2)&=&0,
\end{eqnarray}
\begin{equation}
\delta_{1,5}\equiv |\coth\alpha_{1,5}|-1.
\end{equation}
  These equations are the same as
(\ref{equation-in-the-asymptotic-region-for-X1}) and
(\ref{equation-in-the-asymptotic-region-for-X2}) except for the restoration of
the singularities near $x=0$.
  Note that there is just one additional singularity 
for each of the above two differential equations.
  In the matching region we can neglect the second term in the L.H.S. of these 
equations and the resulting equations have two regular singularities at 
finite $x$.
  These are of analytically solvable form.
  The two independent solutions for $F_{l+1,l}$ are
\begin{eqnarray}
F_{l+1,l}&=& (V_1V_2)^{\frac{1}{4}}
             \zeta_1^{\frac{c}{2}}(1-\zeta_1)^{-\frac{1}{2}}\nonumber \\ 
  & &\cdot(-\zeta_1)^{-a}F(1+a-c,a,a-b+1;\zeta_1^{-1})\nonumber \\
         &=& (V_1V_2)^{\frac{1}{4}}
             \zeta_1^{\frac{c}{2}}(1-\zeta_1)^{-\frac{1}{2}}\cdot
                             \nonumber \\
         & & \left\{F(a,b,c;\zeta_1)
                    \frac{\Gamma(a-b+1)\Gamma(1-c)}
                         {\Gamma(1+a-c)\Gamma(1-b)}
                              \right.\nonumber \\
         & &\;\;\;\;\;+(-\zeta_1)^{1-c}F(1+a-c,1+b-c,2-c;\zeta_1)
                       \cdot\nonumber \\
         & & \left. \;\;\;\;\;\;\;\;\;\;\;\;\;\;\;\;\;\;\;\;\;\;\;\;\;
                    \frac{\Gamma(a-b+1)\Gamma(c-1)}
                         {\Gamma(a)\Gamma(c-b)} \right\}
\end{eqnarray}
\begin{equation}
\zeta_1=-x/\delta_1\;,\;
a=-l-2+\varepsilon(2l+3)\;,\;
b=\varepsilon\;,\;
c=-l+2\varepsilon(l+2)\;,
\end{equation}
and that obtained from the above expression by $a\leftrightarrow
b$.
  The solutions for $G_{l+1,l}$ are obtained from those for
$F_{l+1,l}$ by $\zeta_1\rightarrow\zeta_5=-x/\delta_5$ 
and $l\rightarrow -l-3$.
  Connecting the solutions in the horizon region, the matching region and the
asymptotic region in order gives the value of 
$A$: for $\alpha_n>0$,
\begin{equation}
  A= \frac{\Gamma(\frac{1}{2}-\frac{i\omega}{2\pi T_H})
             \Gamma(l+2)\Gamma(l+2)}
            {\Gamma(\frac{l+2}{2}-\frac{i\omega}{4\pi T_L})
             \Gamma(\frac{l+3}{2}-\frac{i\omega}{4\pi T_R})}
       2^{l+1}(r_0\omega)^{-l-\frac{1}{2}}
       \left(\frac{\delta_5}{\delta_1}\frac{T_L}{T_R}\right)^{\frac{1}{4}},
\end{equation}
and $T_R\leftrightarrow T_L$ for $\alpha_n<0$.
  Matching $F_{l+1,l}$ and $G_{l+1,l}$ give the same result for $A$
in the dilute gas approximation,
but not for $B$. 
  But we do not have to take the mismatch of $B$
very seriously because $|B|\ll |A|$ always holds.

\subsubsection{$\alpha_1<0\;,\;\alpha_5>0$}
  This case is similar to the previous subsubsection and we can obtain the 
result from the one in the previous case by 
$1\leftrightarrow 5$.

\subsection{The greybody factor}
  From the results obtained in the previous subsections, we can calculate
the greybody factor.
  Neglecting $B$ being much smaller than $A$, it becomes as
\begin{equation}
 \sigma_{\rm abs}^{(l+1,l)}
               =\frac{\pi (l+1)(l+2)}{\omega^3}\frac{f_{\rm abs}}{f_{\rm in}}
                 =\frac{\pi^2 r_0^3 (l+1)(l+2)}{r_0\omega}|A|^{-2},
\end{equation}
where the factor $\frac{\pi (l+1)(l+2)}{\omega^3}$ converts the partial-wave 
absorption probability to the absorption cross section of the partial wave.
  Under the dilute gas approximation, it can be written in a simple form:
for $\alpha_n>0$,
\begin{eqnarray}
 \sigma_{\rm abs}^{(l+1,l)}&=& 
    \left(2^{-2}\pi^2 r_0^3 e^{-(\alpha_1+\alpha_5+\alpha_n)}\right)
    \cdot (l+1)(l+2)(r_0\omega)^{2l}\cdot 2^{-2l-2+2q_0} 
                \nonumber \\ & &\cdot
 \left|\frac{\Gamma(\frac{l+1+q_0}{2}-\frac{i\omega}{4\pi T_L})
        \Gamma(\frac{l+2+q_0}{2}-\frac{i\omega}{4\pi T_R})}
       {\Gamma(\frac{1}{2}-\frac{i\omega}{2\pi T_H})
        \Gamma(l+1+\left[\frac{q_0+1}{2}\right])
        \Gamma(l+2+\left[\frac{q_0}{2}\right])}\right|^2,
\label{greybody-factor}
\end{eqnarray}
\[ q_0=\theta(\alpha_1)+\theta(\alpha_5), \]
and $T_R\leftrightarrow T_L$ for $\alpha_n<0$.
  In this expression $[x]$ is the Gauss symbol of $x$ and 
$[n]=[n+1/2]=n$ for integer $n$. 
  Recalling that the area of the outer event horizon is expressed as
\[ {\cal A}_H=2\pi^2 r_0^3 \cosh\alpha_1\cosh\alpha_5\cosh\alpha_n, \]
the first factor in the right hand side of (\ref{greybody-factor})
can be rewritten as
\[2^{-2}\pi^2 r_0^3 e^{-(\alpha_1+\alpha_5+\alpha_n)}
 = \frac{{\cal A}_H e^{|\alpha_1|+|\alpha_5|-\alpha_1-\alpha_5}}
                      {1+e^{2\alpha_n}} \]
and approaches to ${\cal A}_H$ when $\alpha_{1,5}$ are negative
and $\alpha_n$ is negatively large.
  If either or both of $\alpha_{1,5}$ are positive, the greybody
factor becomes small and the emission from this fermion mode is
suppressed.
  This is because the fermion mode we have been considering has
negative D1,D5 and KK charges.

  In the following we shall consider the case of $\alpha_{1,5,n}<0$.
  In this case the greybody factor becomes
\begin{eqnarray}
 \sigma_{\rm abs}^{(l+1,l)}&=&  
    \frac{{\cal A}_H}{1+e^{2\alpha_n}}
    \cdot \frac{(l+1)(l+2)}{4\left\{l!(l+1)!\right\}^2}
         \left(\frac{r_0\omega}{2}\right)^{2l} \nonumber \\ & &\;\;\;
 \cdot\left|\frac{\Gamma(\frac{l+1}{2}-\frac{i\omega}{4\pi T_R})
        \Gamma(\frac{l+2}{2}-\frac{i\omega}{4\pi T_L})}
       {\Gamma(\frac{1}{2}-\frac{i\omega}{2\pi T_H})}\right|^2.
\end{eqnarray}
  The form of 
$\sigma_{\rm abs}$ is strikingly analogous to the one for scalar field
computed in \cite{9701187},\cite{9702015}.
  We expect that the same expression can be obtained from the
effective string theory calculation as the collision process of R- and 
L- movers emitting a fermion into spacetime.
  The relevant interaction vertex must be a three-point coupling of
spacetime fermion and one holomorphic and one antiholomorphic
operators, each with weight $\frac{l+1}{2}$ and $\frac{l+2}{2}$,
respectively.

  In the case of s-wave ($l=0$), $\sigma_{\rm abs}^{(1,0)}$ becomes
\begin{equation}
  \sigma_{\rm abs}^{(1,0)}= \frac{{\cal A}_H}{1+e^{2\alpha_n}}
\frac{e^{\frac{\omega}{T_H}}+1}
     {(e^{\frac{\omega}{2T_R}}+1)(e^{\frac{\omega}{2T_L}}-1)}
\frac{\omega}{4T_L}. 
\end{equation}
  Sending $\alpha_n$ to a large negative value, it becomes as
\begin{equation}
  \sigma_{\rm abs}^{(1,0)}=\frac{{\cal A}_H}{2}.
\end{equation}
  The decay rate is expressed in general case as
\begin{equation}
\Gamma(\omega)^{(l+1,l)}=\sum_s \frac{d^4k}{(2\pi)^4}
        \frac{\sigma_{\rm abs}^{(l+1,l)}}{e^{\frac{\omega}{T_H}}+1}
\end{equation}
and, in the s-wave case and for a large negative $\alpha_n$ 
we obtain the Hawking formula by summing over polarizations.
\begin{equation}
\Gamma(\omega)^{(1,0)}=\frac{d^4k}{(2\pi)^4}
        \frac{{\cal A}_H}{e^{\frac{\omega}{T_H}}+1}.
\end{equation}

\section{Conclusion}
\setcounter{equation}{0}
  We would like to conclude this paper by reviewing the calculation
and summarizing the results.
  We considered the emission of spin-$\frac{1}{2}$ particles from
charged, nonrotating black holes in five-dimensional $N=8$
supergravity.
  The Dirac fields in this theory have a `Pauli term' or a nonminimal
coupling with $U(1)$ field strengths and through this coupling the
emitted Dirac particles can carry the charges out from the black hole.
  This interaction term can be diagonalized for a certain subset of spinor 
fields which are interpreted, from the type IIB on $T^5$ viewpoint,
as gravitini whose vector indices are those of internal $T^4$. 
  We derived the radial wave equation and solved it using the similar
method as in the case of scalar field.
  The wave equation turned out to have a $SL(2,R)\times SL(2,R)$
symmetry at the horizon region, as in the case of scalar field and this
property is considered to be crucial for the `effective string'
interpretation.
  The problem of matching the solutions was much more 
complicated than the scalar case.
  We finally obtained the absorption cross section or the greybody
factor.
  The expression is very similar to the one for scalar field and we expect
that it can be re-derived from the effective string theory calculation.
  In the near-extremal limit the greybody factor approaches to the 
horizon area times an unusual factor $1/2$.
  This is canceled by another factor 2 that arises from summing over
spin degrees of freedom of the emitted fermion and the Hawking formula
are obtained.

  Let us note that, as we said in the previous section, the
enhancement or suppression of the emission according to the charges of the 
black hole is due to the charges of the emitted fermion.
  The fermion mode we have been considering is
the fermion with negative KK,D1,D5 charges.
  Actually we have discarded the seven alternatives, namely, the
fermion modes with the other signs of KK,D1,D5 charges by choosing 
$\epsilon_1=\epsilon_2=-1$ in (\ref{the-Dirac-equation-diagonalized}) 
and taking $(F_{l+1,l},G_{l+1,l})$ rather than $(F_{l,l+1},G_{l,l+1})$.
  Unless $\alpha_{1,5,n}$ are all negative the emission from the mode
we have considered are suppressed, but there emerges another mode whose 
emission are enhanced.

  We are not sure if the greybody factor (\ref{greybody-factor}) can
be fully explained, including the overall numerical factor,
by the effective string theory, though it is strongly expected.
  To see the full coincidence with the effective string theory calculation,
we need more detailed knowledge about its worldsheet action, in
particular its fermionic degrees of freedom and their interaction with
spacetime fields.
  This will be our future problem.
  
\bigskip

\noindent{\large \bf Acknowledgment}

  The author thanks T.Eguchi for introduction to this subject and
helpful discussions, S.R.Das for communications and valuable comments.
  The author is also grateful to A.Matsuda and M.Matsuda
for discussions.

\newpage
\appendix

\section{The properties of $\Psi^{(l)}_{jj'mm'}$}
\setcounter{equation}{0}
  In this appendix we list some basic properties of $\Psi^{(l)}_{jj'mm'}$.

  $\Psi^{(l)}_{jj'mm'}$ has the following eigenvalues:
\begin{eqnarray} 
  J^2&=&4j(j+1)+4j'(j'+1),\nonumber \\
  L^2&=&2l(l+2)\nonumber \\
  \Sigma^2&=&3,\nonumber \\
  \Sigma\cdot L&=&\left\{\begin{array}{lll}
          l& {\rm  for}& \Psi^{(l)}_{\frac{l+1}{2}\frac{l}{2}mm'},
                         \Psi^{(l)}_{\frac{l}{2}\frac{l+1}{2}mm'} \\
       -l-2& {\rm  for}& \Psi^{(l)}_{\frac{l-1}{2}\frac{l}{2}mm'},
                         \Psi^{(l)}_{\frac{l}{2}\frac{l-1}{2}mm'}
                         \end{array} \right.
\end{eqnarray}

  The operator $\Pi\equiv\frac{1}{r}(\Gamma\cdot x)$ commutes with 
$J_{ij}$ and $\Pi^2=1$.
  $\Pi$ shifts the value of $l$ by $\pm 1$.
\begin{eqnarray}
    \Pi\Psi^{(l)}_{\frac{l\pm 1}{2}\frac{l}{2}mm'}
   &=&\Psi^{(l\pm 1)}_{\frac{l\pm 1}{2}\frac{l}{2}mm'} \nonumber \\ 
    \Pi\Psi^{(l)}_{\frac{l}{2}\frac{l\pm 1}{2}mm'}
   &=&\Psi^{(l\pm 1)}_{\frac{l}{2}\frac{l\pm 1}{2}mm'}.
\end{eqnarray}

  $\Gamma^0$ commutes with $J_{ij}$ and the Hamiltonian.
  $(\Gamma^0)^2=-1$ and we set its phase as
\begin{eqnarray}
  \Gamma^0\Psi^{(l)}_{\frac{l\pm 1}{2}\frac{l}{2}mm'}
    &=& -i\Psi^{(l)}_{\frac{l\pm 1}{2}\frac{l}{2}mm'}  \nonumber \\
  \Gamma^0\Psi^{(l)}_{\frac{l}{2}\frac{l\pm 1}{2}mm'}
    &=&  i\Psi^{(l)}_{\frac{l}{2}\frac{l\pm 1}{2}mm'}.
\end{eqnarray}

  The orthonormality of $\Psi^{(l)}_{jj'mm'}$ is set as
\begin{equation}
  \int d\Omega_{(3)}\Psi^{\dagger(l)}_{j_1 j_2m_1m_2}
                    \Psi^{(l')}_{j'_1 j'_2m'_1m'_2}
  =\delta_{ll'}\delta_{j_1j'_1}\delta_{j_2j'_2}\delta_{m_1m'_1}
   \delta_{m_2m'_2}. \nonumber
\end{equation}

  We finally list some identities:
\begin{eqnarray}
  (\Gamma\cdot x)(\Gamma\cdot\partial)&=&
  r\frac{d}{dr}-(\Sigma\cdot L) \\
  (\Gamma\cdot\partial)(\Gamma\cdot x)&=&
  r\frac{d}{dr}+4+(\Sigma\cdot L) \\
  \Pi(\Sigma\cdot L)+(\Sigma\cdot L)\Pi &=& -3\Pi.
\end{eqnarray} 

\section{The relation between the plane wave and the spherical wave}
\setcounter{equation}{0}
  In this appendix we shall review the plane wave and the partial wave 
solutions of the Klein-Gordon and the Dirac equations in a flat 
five-dimensional spacetime.

  We begin with the Klein-Gordon equation.
  The plane wave solution is expressed as
\begin{equation}
 \phi_k(x)=e^{ik\cdot x}
\end{equation}
and its orthonormality are expressed as
\begin{equation}
  \int dx^4 \phi^\ast_k(x)\cdot\phi_{k'}(x)=(2\pi)^4\delta^4(k-k').
\end{equation}

  The spherical-wave solution can be written as
\begin{equation}
  \phi_{[\omega,l,m_1,m_2]}(x)=\sqrt{2\pi\omega^3}\cdot(\omega r)^{-1}
  J_{l+1}(\omega r)Y^{(l)}_{m_1m_2}
\end{equation}
where $Y^{(l)}_{m_1m_2}$ are the four-dimensional spherical harmonics and 
the indices represent their eigenvalues of the orbital angular momentum:
\begin{eqnarray}
  L_{ij}L_{ij}\cdot Y^{(l)}_{m_1m_2}&=& 2l(l+2)Y^{(l)}_{m_1m_2} \nonumber \\
  \frac{1}{2}(L_{12}+L_{34})Y^{(l)}_{m_1m_2}&=&m_1Y^{(l)}_{m_1m_2}\nonumber\\
  \frac{1}{2}(L_{12}-L_{34})Y^{(l)}_{m_1m_2}&=&m_2Y^{(l)}_{m_1m_2}.
\end{eqnarray}
  Its orthonormality is expressed as
\begin{equation}
  \int dx^4 \phi^\ast_{[\omega,l,m_1,m_2]}(x)
                 \phi_{[\omega',l',m'_1,m'_2]}(x)
 = (2\pi)\delta(\omega-\omega')\delta_{ll'}\delta_{m_1m'_1}\delta_{m_2m'_2}
\end{equation}

  We consider the Dirac equation next.
  We are going to work in the representation where
\begin{equation}
2\Sigma_{12}= \left(\begin{array}{cccc}
                1     & \cdot & \cdot & \cdot \\
                \cdot & -1    & \cdot & \cdot \\ 
                \cdot & \cdot & 1     & \cdot \\
                \cdot & \cdot & \cdot & -1    \end{array}\right)\;\;,\;\;
2\Sigma_{34}= \left(\begin{array}{cccc}
                1     & \cdot & \cdot & \cdot \\
                \cdot & 1     & \cdot & \cdot \\ 
                \cdot & \cdot & -1    & \cdot \\
                \cdot & \cdot & \cdot & -1    \end{array}\right)\nonumber 
\end{equation}
\begin{equation}
\Gamma_1    = \left(\begin{array}{cccc}
                \cdot & -i    & \cdot & \cdot \\
                i     & \cdot & \cdot & \cdot \\ 
                \cdot & \cdot & \cdot & i     \\
                \cdot & \cdot & -i    & \cdot  \end{array}\right)\;\;,\;\;
\Gamma^0=-\Gamma_0\
            = \left(\begin{array}{cccc}
                -i    & \cdot & \cdot & \cdot \\
                \cdot & i     & \cdot & \cdot \\ 
                \cdot & \cdot & i     & \cdot \\
                \cdot & \cdot & \cdot & -i    \end{array}\right)
\end{equation}
where $\Sigma_{ij}=-\frac{i}{4}[\Gamma_i,\Gamma_j]$. 
  The normalized massless plane wave solutions are obtained as 
the four column vectors in the $(4\times 4)$matrix
\begin{equation}
 \frac{1}{\sqrt{2}k^0}(i\Gamma^m\partial_m)e^{ik\cdot x}.
\end{equation}
  Note that only two of the four column vectors are linearly independent.
  Denoting the two independent solutions with the given momentum $k$ as
$\psi_{k,1}$ and $\psi_{k,2}$, the orthonormality is expressed as
\begin{equation}
  \int d^4x \psi^\dagger_{k,s}(x)\psi_{k',s'}(x)=(2\pi)^4\delta^4(k-k')
       \delta_{ss'}.
\end{equation}
  When the momentum is along the $x_1$-axis the two linearly independent
solutions are written as
\begin{equation}
\psi_{k_x,+}(x) = \frac{1}{\sqrt{2}}
         \left(\begin{array}{c}1\\1\\ \cdot\\ \cdot\end{array}\right)
         e^{-ik^0(t-x_1)} \;\; , \;\;
\psi_{k_x,-}(x) = \frac{1}{\sqrt{2}}
         \left(\begin{array}{c}\cdot\\ \cdot\\1\\1\end{array}\right)
         e^{-ik^0(t-x_1)}
\end{equation}
where the indices $\pm$ represent the eigenvalue $\pm 1$ of $2\Sigma_{34}$.
  The spherical wave solution is written as
\begin{eqnarray}
\lefteqn{\psi_{[\omega,\frac{l+1}{2}\frac{l}{2}mm']}(x)} \nonumber \\
         &=& \frac{1}{\sqrt{2}}e^{-i\omega t}\left[
         \frac{\sqrt{2\pi\omega^3}}{\omega r}J_{l+1}(\omega r)
         \Psi^{(l)}_{\frac{l+1}{2}\frac{l}{2}mm'}
        +\frac{\sqrt{2\pi\omega^3}}{\omega r}J_{l+2}(\omega r)
         \Psi^{(l+1)}_{\frac{l+1}{2}\frac{l}{2}mm'} \right] \nonumber \\
\lefteqn{\psi_{[\omega,\frac{l}{2}\frac{l+1}{2}mm']}(x)} \nonumber \\
         &=& \frac{1}{\sqrt{2}}e^{-i\omega t}\left[
         \frac{\sqrt{2\pi\omega^3}}{\omega r}J_{l+1}(\omega r)
         \Psi^{(l)}_{\frac{l}{2}\frac{l+1}{2}mm'}
        -\frac{\sqrt{2\pi\omega^3}}{\omega r}J_{l+2}(\omega r)
         \Psi^{(l+1)}_{\frac{l}{2}\frac{l+1}{2}mm'} \right] \nonumber \\
 & &
\end{eqnarray}
where $\Psi^{(l)}_{jj'mm'}$ are the eigenspinors of the total angular momentum
and they are expressed, in our representation of $\Gamma$ matrix, as
\[
\Psi^{(l)}_{\frac{l+1}{2}\frac{l}{2}mm'}=
 \left( \begin{array}{c}
        \sqrt{\frac{l+2m+1}{2l+2}}Y^{(l)}_{m-\frac{1}{2},m'}\\
        0 \\0\\
        \sqrt{\frac{l-2m+1}{2l+2}}Y^{(l)}_{m+\frac{1}{2},m'} \end{array}\right)
   \;,\;
\Psi^{(l+1)}_{\frac{l+1}{2}\frac{l}{2}mm'}=
 \left( \begin{array}{c}
        0\\-\sqrt{\frac{l+2m'+2}{2l+4}}Y^{(l+1)}_{m,m'+\frac{1}{2}}\\
        \sqrt{\frac{l-2m'+2}{2l+4}}Y^{(l+1)}_{m,m'-\frac{1}{2}}\\0 
        \end{array} \right),
\]
\[
\Psi^{(l)}_{\frac{l}{2}\frac{l+1}{2}mm'}=
 \left( \begin{array}{c}
        0\\ \sqrt{\frac{l-2m'+1}{2l+2}}Y^{(l)}_{m,m'+\frac{1}{2}}\\
        \sqrt{\frac{l+2m'+1}{2l+2}}Y^{(l)}_{m,m'-\frac{1}{2}}\\0
         \end{array} \right)
   \;,\;
\Psi^{(l+1)}_{\frac{l}{2}\frac{l+1}{2}mm'}=
 \left( \begin{array}{c}
        \sqrt{\frac{l-2m+2}{2l+4}}Y^{(l+1)}_{m-\frac{1}{2},m'}\\0\\0\\
       -\sqrt{\frac{l+2m+2}{2l+4}}Y^{(l+1)}_{m+\frac{1}{2},m'} \end{array}
        \right).
\]
\begin{eqnarray}
  & &
\end{eqnarray}

  Finally we would like to calculate the geometric factor appearing in 
(\ref{greybody-factor}).
  It is calculated as
\[ \sum_{m,m'}|f_{\frac{l+1}{2},\frac{l}{2},mm'}|^2 \]
where the coefficients $f_{\frac{l+1}{2},\frac{l}{2},mm'}$ are determined
through
\begin{equation}
 \int d^4x \psi^\dagger_{[\omega,\frac{l+1}{2},\frac{l}{2},mm']}(x)
       \psi_{k_x,+}(x)
   = (2\pi)\delta(\omega-k^0)\cdot f_{\frac{l+1}{2},\frac{l}{2},mm'}.
\end{equation}
  We define the polar coordinates $r,\theta,\varphi_1,\varphi_2$ as
\begin{eqnarray}
\lefteqn{(x_1,x_2,x_3,x_4)=} \nonumber  \\
 & &(r\cos\theta\cos\varphi_1,r\cos\theta\sin\varphi_1,
     r\sin\theta\cos\varphi_2,r\sin\theta\sin\varphi_2).
\end{eqnarray}
  In this coordinate system the spherical harmonics have the form
\begin{equation}
  Y^{(l)}_{mm'}(\theta,\varphi_1,\varphi_2)
  =y^{(l)}_{mm'}(\theta)e^{i(m+m')\varphi_1+i(m-m')\varphi_2}.
\end{equation}
  Recalling the decomposition formula
\begin{equation}
  e^{i\omega r\cos \theta}=\sum_n i^n (n+1)\sqrt{\frac{4\pi}{\omega^3}}\cdot
  \frac{\sqrt{2\pi\omega^3}}{\omega r}J_{n+1}(\omega r)\cdot 
  X^{(n)}(\cos\theta)
\end{equation}
\begin{eqnarray}
  X^{(n)}(\cos\theta)&=& \frac{\sin\left\{(n+1)\theta\right\}}
                              {\sqrt{2\pi^2}\sin\theta} \nonumber \\
  \int d\Omega_{(3)} |X^{(n)}|^2&=& 1
\end{eqnarray}
we can calculate the coefficient $f_{\frac{l+1}{2},\frac{l}{2},mm'}$ by
\begin{eqnarray}
  f_{\frac{l+1}{2},\frac{l}{2},mm'}&=&
 i^l\sqrt{\frac{\pi}{2\omega^3}}\int d\Omega_{(3)}\cdot
   \nonumber \\ & &\;\;\;
  \left[\sqrt{(l+2m +1)(l+1)}Y^{(l  )\ast}_{m-\frac{1}{2},m'}
        X^{(l)}(\cos\theta\cos\varphi_1)\right.
    \\ & &\;\left.
      -i\sqrt{(l+2m'+2)(l+2)}Y^{(l+1)\ast}_{m,m'+\frac{1}{2}}
        X^{(l+1)}(\cos\theta\cos\varphi_1)\right]. \nonumber 
\end{eqnarray}
  Taking the reality of $X^{(l)}$ and the fact that
\[ \int d\Omega_{(3)}Y^{(l)\ast}_{m,m'}X^{(l)}(\cos\theta\cos\varphi_1)=
   \int d\Omega_{(3)}Y^{(l)\ast}_{-m,-m'}X^{(l)}(\cos\theta\cos\varphi_1) \]
into consideration, the calculation becomes very simple and we obtain
\begin{equation}
\sum_{m,m'}|f_{\frac{l+1}{2},\frac{l}{2},mm'}|^2
  =\frac{\pi}{\omega^3}(l+1)(l+2).
\end{equation}

\end{document}